\documentclass[preprint,3p,times,onecolumn]{elsarticle}
\usepackage{amsmath,amssymb,multirow,graphicx,pennames,fixltx2e}

\journal{Elsevier}

\usepackage{hyperref}

\begin{document}

\begin{frontmatter}

\title{Improved primary vertex finding for collider detectors}

\author{Ferenc Sikl\'er}
\ead{sikler@rmki.kfki.hu}
\address{KFKI Research Institute for Particle and Nuclear Physics, Budapest,
Hungary}

\begin{abstract}

Primary vertex finding for collider experiments is studied. The efficiency and
precision of finding interaction vertices can be improved by advanced
clustering and classification methods, such as agglomerative clustering with
fast pairwise nearest neighbor search, followed by Gaussian mixture model or
k-means clustering.

\end{abstract}


\end{frontmatter}

\section{Introduction}
\label{sec:intro}

In this paper one-dimensional vertex finding for minimum bias proton-proton
collisions is studied, with a special care for detecting {\it all} inelastic
interactions, even in high pile-up. For the procedure clean primary particles
are needed with good estimate of their starting longitudinal
coordinate $z$, at the point of closest approach to beam line, as well as
its standard variation $\sigma_z$. The selection of particles is
often based on the value of the impact parameter $d$ of the track and its
estimated standard deviation $\sigma_d$, e.g. requiring $d < 3 \sigma_d$ (beam
spot constraint).

Finding the vertex or vertices of inelastic interactions in a collider is
essential for physics analyses. The aim is to identify primary vertices
efficiently with low background. Several methods and algorithms have been
developed in the past.

One of them \cite{Badala:2001em} works well for single, high multiplicity
events (e.g. heavy-ion collisions). It first estimates the vertex position by
the centroid of $z$ coordinates of hits left in the detector. It makes
use of the correlation between the hits in separate layers and estimates the
location of the vertex.

Another one \cite{Costa:2007zzc} takes all reconstructed tracks which are
compatible with the beam spot. The mode of their $z$ coordinates is used as
vertex candidate. Tracks not compatible with the vertex are discarded,
the rest is used for an adaptive multi-vertex fit. The remaining tracks are
used to find a new vertex candidate. The procedure continues until a minimum of
two tracks compatible with a common vertex is reached. The method has very high
efficiency to find the correct vertex in a sample of $t\overline{t}$ events.

\begin{figure*}

 \begin{center}
  \input{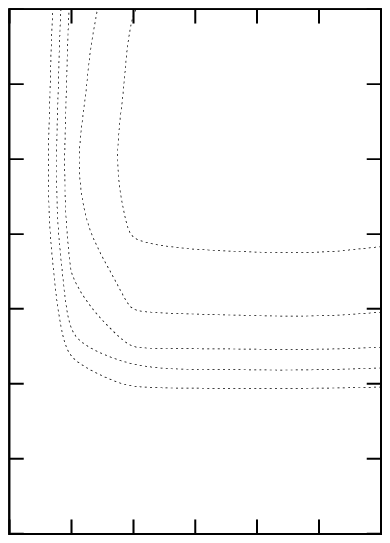}
  \input{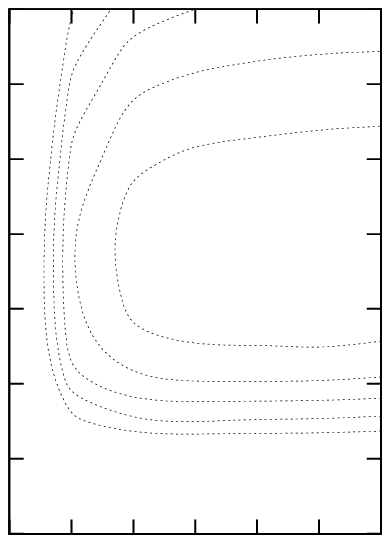}
  \input{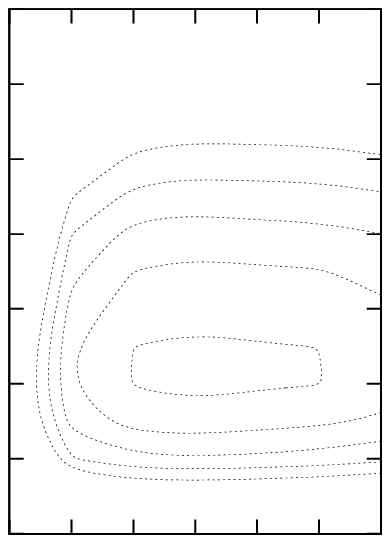}
 \end{center}

 \caption{Optimization of the standard method for $n_{min}$ = 2 for several
$K_{sim}$ values. The contour lines of $X^2$ are drawn, the place (+) and
value of the minimum are indicated.}

 \label{fig:min2}

\end{figure*}

This article is organized as follows: Sec.~\ref{sec:standard} introduces a
standard method of primary vertex finding, defines some performance measures
and discusses its optimization. Sec.~\ref{sec:agglo} shows the possible
application of hierarchical clustering, while Sec.~\ref{sec:class} deals with
more sophisticated methods, such as Gaussian mixture models and k-means
clustering. The details of the Monte Carlo simulation are given
in Sec.~\ref{sec:simulation}. The comparisons of performance and timing are
shown in Sec.~\ref{sec:result}. This work ends with conclusions.

\section{The standard method}
\label{sec:standard}

In a third method \cite{Adam:2006ki} primary vertex candidates are obtained by
clustering selected tracks according to their $z$ coordinate. The so called
divisive method looks for large intervals without tracks and divides the $z$
axis into several regions. For each region the vertex position is computed with
a weighted average of all compatible tracks. Tracks not compatible with that
average vertex position are discarded, but they are recovered to make a new
vertex candidate.

First, the tracks are ordered according to increasing $z$ value. The ordered
list is scanned to form a cluster until at least $n_{min}$ consecutive tracks
separated by more than $z_{sep}$ are found, at which point another cluster is
built. The position of each of these primary vertex clusters is determined
iteratively. A cleaning procedure is applied, rejecting the tracks farther from
the estimated primary vertex position $z_k$ than $n_\sigma$ standard
deviations ($|z - z_k | \le n_\sigma \cdot \sigma_z$). The positions of the primary
vertices are recomputed with the remaining tracks. The procedure iterated until
each remaining track is compatible with its associated cluster according to the
above criterion.

For specific signal events the settings $n_\sigma$ = 5, $z_{sep}$ = 50 -- 500
$\mu$m and $n_{min}$ = 2 appeared to be most successful.  The efficiency for
reconstructing and correctly tagging the primary vertex of the signal event
depends on the number of charged tracks produced at the vertex, and for low
luminosity it ranges from 76\% to to 99\% depending on the signal process
studied.

In the following comparisons the above mentioned method will be used as
standard vertex finder since it is best suited for finding multiple vertices
opposed to others (Sec.~\ref{sec:intro}) which search only for the signal
vertex.

\subsection{Measures of performance}

\def\SV{simulated vertex}
\def\SVs{simulated vertices}
\def\RV{reconstructed vertex}
\def\RVs{reconstructed vertices}

 A vertex is associated to another vertex if more than half of its tracks are
shared. This way a reconstructed (simulated) vertex can be associated to a
simulated (reconstructed) vertex: the number of possible associations is zero
or one.  A \SV\ is reconstructed $n$ times if there are $n$ \RVs\ which are
associated to it.  The {\it efficiency} gives the fraction of \SVs\ which are
at least once reconstructed. The {\it lost} vertex rate shows the fraction of
\SVs\ which are not reconstructed. The {\it split} vertex rate shows the
fraction of \SVs\ which are more than once reconstructed.  A \RV\ is a {\it
fake} if it has no associated \SVs. A \RV\ is {\it merged} if it has more than
one associated \SVs.  These quantities can be studied as a function of fixed or
Poisson distributed event-by-event pile-up.

\begin{figure*}
 
 \begin{center}
  \input{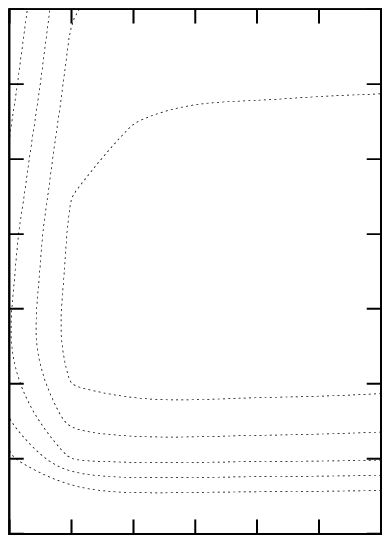}
  \input{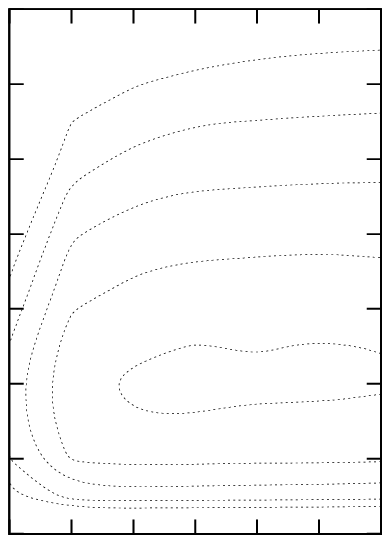}
  \input{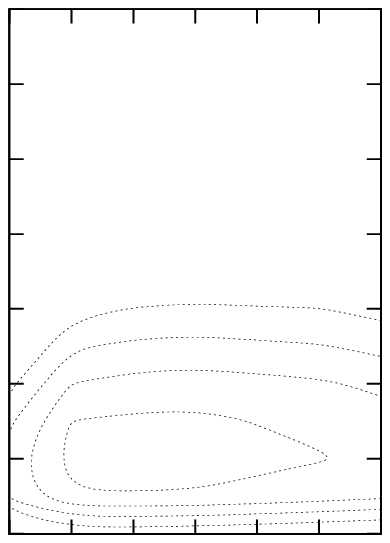}
 \end{center}

 \caption{Optimization of the standard method for $n_{min}$ = 3 for several
$K_{sim}$ values. The contour lines of $X^2$ are drawn, the place (+) and
value of the minimum are indicated.}

 \label{fig:min3}
\end{figure*}

\subsection{Optimization of the standard method}

The standard algorithm can be optimized by choosing $n_{min}$, $n_\sigma$
and $z_{sep}$ such that the performance of vertex finding is best. As it was
presented in Sec.~\ref{sec:standard} the roles of the parameters are

\begin{itemize}
 \item $n_{min}$: minimum number of tracks required to form a cluster. In order
to have low bias, it is reasonable to look at settings $n_{min} =$ 2 and 3.
 \item $n_\sigma$: a track is compatible with a vertex if it is closer than
$n_\sigma$ times the estimated standard deviation of $z$.
 \item $z_{sep}$: maximum distance between two adjacent tracks that belong
       to the same initial cluster.
\end{itemize}

The measure of goodness, the merit function $X^2$ to minimize, can be chosen
as the sum of the fractions of lost and split simulated vertices
and that of the fake and merged reconstructed vertices:
\begin{equation*}
 X^2 = \biggl\langle \frac{K_{lost} + K_{multi}}{K_{sim}} +
                     \frac{K_{fake} + K_{merged}}{K_{rec}} \biggr\rangle.
\end{equation*}

The $X^2$ values for several settings, using $10^4$ inelastic proton-proton
events, were calculated on a grid: $K_{sim}$ = 1, 2, 4 and 8; $n_\sigma$ = 2,
2.5, \dots, 5; $z_{sep}$ = 0.1, 0.15, \dots, 0.8 cm; $n_{min}$ = 2, 3. The
obtained contour lines and the places of the minima are shown in
Fig.~\ref{fig:min2} ($n_{min}$=2) and Fig.~\ref{fig:min3} ($n_{min}$=3). The
best values are also given in Table~\ref{tab:opt}. Both the optimal $n_\sigma$
and $z_{sep}$ depend on the real vertex multiplicity $K_{sim}$. In case of a
single simulated vertex a big $n_\sigma$ and $z_{sep}$ is needed for good
reconstruction. It turns out that $n_{min}$=3 gives comparable or better
performance for all $K_{sim}$ values than those with $n_{min}$=2. The setting
\begin{align*}
 n_{min}  &= 3, &
 n_\sigma &= 3.0, &
 z_{sep}  &= 0.3~\mathrm{cm}
\end{align*}

\noindent is an acceptable compromise. Hence for the standard method these
values were used in the following studies.

\begin{table}

 \caption{Optimized parameters for the standard method. Best $n_\sigma$,
$z_{sep}$ and minimal $X^2$ values for settings $n_{min}$ = 2 and 3
using 1, 2, 4 or 8 simulated vertices.}

 \label{tab:opt}
 
 \begin{center}
  \begin{tabular}{ccccc}
  \hline
  $n_{min}$ & $K_{sim}$ & $n_\sigma$ & $z_{sep}$ [cm] & $X^2_{min}$ \\
  \hline
  \multirow{4}{*}{2} 
    & 1 & big & big  & small \\
    & 2 & 5.0 & 0.65 & 0.113 \\
    & 4 & 4.0 & 0.45 & 0.261 \\
    & 8 & 3.5 & 0.30 & 0.462 \\
  \hline
  \multirow{4}{*}{3} 
    & 1 & big & big  & small \\
    & 2 & 3.5 & 0.40 & 0.120 \\
    & 4 & 3.5 & 0.30 & 0.225 \\
    & 8 & 3.0 & 0.20 & 0.382 \\
  \hline
  \end{tabular}
 \end{center}

\end{table}


\section{Hierarchical clustering}

\label{sec:agglo}

Since the track multiplicities of vertices greatly vary, for efficient
classification (Sec.~\ref{sec:class}) the precise and unbiased designation of
cluster centers is essential. The most widely used methods for hierarchical
clustering are agglomerative methods \cite{Press:1058313}. They start by
connecting individual data points into small clusters, then connect those
clusters, and so forth.  The pairwise nearest neighbor method (PNN) using
weighted averages proved to be a highly successful method for agglomerative
clustering. In this work an advanced implementation, the fast pairwise nearest
neighbor method (fPNN) \cite{Franti:2000}, was used where the list of closest
neighbors is kept and updated.

The distance $d$ of two tracks $i$ and $j$ is defined as
\begin{equation*}
 d^2 = \frac{(z_i - z_j)^2}{\sigma_i^2 + \sigma_j^2}.
\end{equation*}

\noindent Note that $d$ is essentially the log-likelihood of the event that the two tracks
belong to the same vertex. The clustering starts with all reconstructed
tracks, each of them being a cluster with a single track. At each step the
clusters with the smallest distance are found and the two clusters are joined:
\begin{align*}
 z        &= \frac{z_i/\sigma_i^2 + z_j/\sigma_j^2}
                  {  1/\sigma_i^2 +   1/\sigma_j^2}, &
 \sigma^2 &= \frac{1}
                  {  1/\sigma_i^2 +   1/\sigma_j^2}.
\end{align*}

\noindent The procedure is repeated until only some $K$ clusters remain and the
resulting vertex positions are passed to other classification methods for
further refinement. More on the selection of optimal $K$ value is given in
Sec.~\ref{sec:class}. 

\begin{figure}

 \begin{center}
  \input{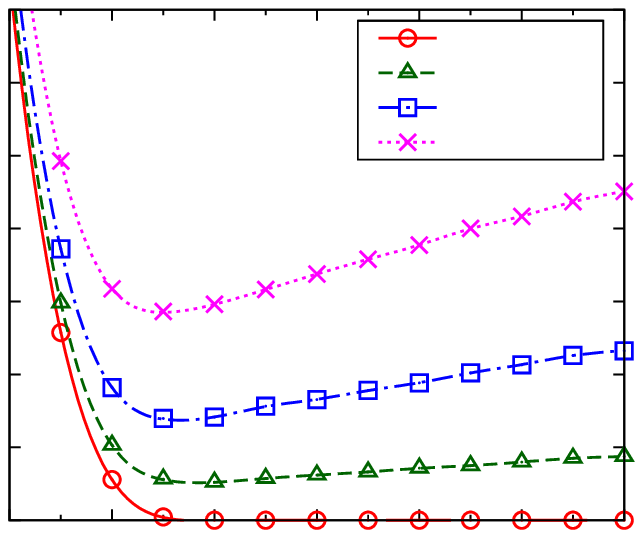}
 \end{center}

 \caption{Optimization of the agglomerative clustering method for several
simulated vertex multiplicities. The lines are drawn to guide the eye.}
  
 \label{fig:tree}

\end{figure}

The $K$ value can also be chosen, and the clustering can also be stopped, if at
a step the smallest distance gets bigger than a given number $d_{max}$. The
merit value $X^2$ as function of $d_{max}$ for several simulated vertex
multiplicities $K_{sim} =$ 1, 2, 4 and 8 is shown in Fig.~\ref{fig:tree}. The
choice of $d_{min} \approx $ 8 appeared to give the best result. For performance
plots of this method see Sec.~\ref{sec:result} (labeled with fPNN).

Another agglomerative clustering method, neighbor
joining, was also tested but provided much poorer results and it was not
studied further.

\section{Classification with unsupervised learning}

\label{sec:class}

\subsection{Gaussian mixture model}

\label{sec:gaussm}

Gaussian mixture models are examples of classification by unsupervised learning
where the solution is achieved by series of expectation and maximization steps
\cite{Press:1058313}.  We have $N$ tracks with estimated longitudinal
coordinates $z_n$ at their closest approach to the beam-line and their expected
standard deviations $\sigma_n$ ($n=1,2,\dots,N$). If the number of interaction
vertices $K$ in a bunch crossing is {\it given}, the task is to find the means of the
longitudinal coordinates $\hat{z}_k$ and their weights, fraction of attached to
all tracks, $\hat{P}(k)$ of each vertex ($k = 1,2,\dots,K$).

The likelihood of finding the tracks at positions $z_n$ is a product
\begin{gather}
 {\cal L} = \prod_n P(z_n)
 \label{eq:likelihood}
\intertext{or a sum}
 \chi^2 = -2 \log {\cal L} = -2 \sum_n \log P(z_n)
 \label{eq:loglikelihood}
\end{gather}

\noindent where $P(z_n)$ is the so called the mixture weight for track $n$. It
can be split into contributions from each vertex
\begin{gather*}
  P(z_n) = \sum_k P(z_n,\sigma_n | \hat{z}_k) \hat{P}(k)
\intertext{where the conditional probability is a Gaussian}
  P(z_n,\sigma_n  |\hat{z}_k) =
   \frac{1}{\sigma_n \sqrt{2\pi}}
    \exp\left[-\frac{(z_n - \hat{z}_k)^2}{2\sigma_n^2}\right].
\end{gather*}

\noindent
The probability that track $n$ came from interaction vertex $k$ is
\begin{equation*}
 p_{nk} = \frac{P(z_n,\sigma_n | \hat{z}_k) \hat{P}(k)} {P(z_n)}
\end{equation*}

\noindent where $p$ is also known as the responsibility matrix. In summary, in
the expectation step $p$ can be calculated for given values of means
$\hat{z}_k$ and weights $\hat{P}(k)$ for all vertices. The procedure starts by
using the results of agglomerative clustering for $K$ vertices
(Sec.~\ref{sec:agglo}). During the maximization step, these means and weights
are estimated from $p$ as
\begin{align*}
 \hat{z}_k  &= \frac{\sum_n p_{nk} z_n}{\sum_n p_{nk}} &
 \hat{P}(k) &= \frac{\sum_n p_{nk}}{N}.
\end{align*}

The expectation step followed by the minimization step will increase the
likelihood value, or decrease $\chi^2$. Thus repeated iterations will converge
to an extremum. In practice the process can be stopped if $\chi^2$ decreased
only by a small amount or the number of iterations exceeded some limit. For
performance plots of this method see Sec.~\ref{sec:result} (labeled with
fPNN+GaussM).

\begin{figure*}
 
 \begin{center}
  \input{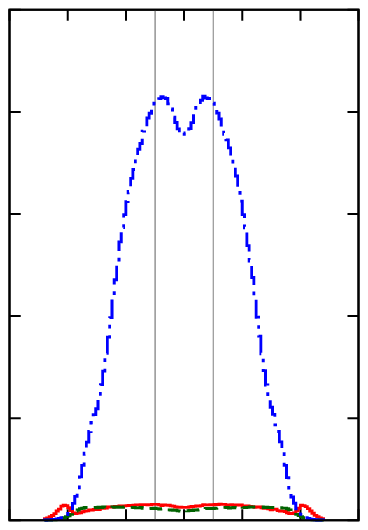}
  \input{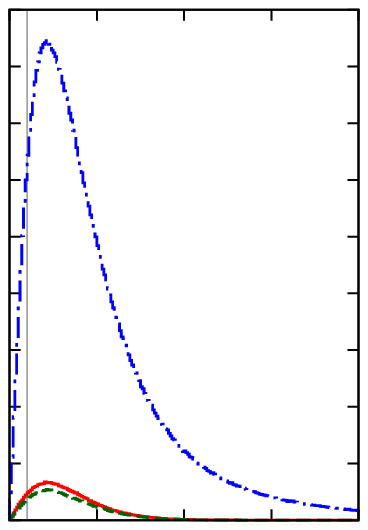}
  \input{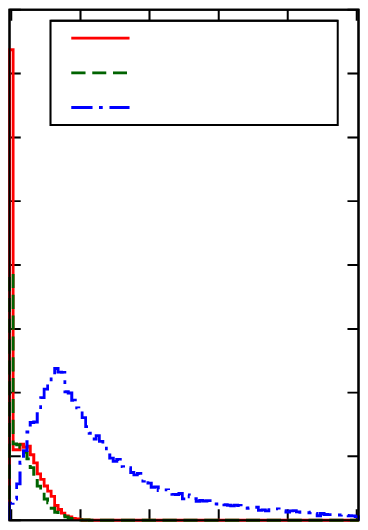}
 \end{center}

 \caption{Left and center: Pseudo-rapidity and transverse momentum
distributions of produced charged particles from inelastic proton-proton
collisions at $\sqrt{s} =$ 10~TeV according to Pythia 6.4. The assumed
acceptance window ($|\eta| <$ 2.5 and $p_T <$ 0.1 GeV/$c$) is indicated by the
grey vertical lines. The event-by-event number distribution of accepted
particles are shown right. The three curves correspond to contributions from
single-, double diffractive and non-diffractive processes.}

 \label{fig:sim}

\end{figure*}

\subsection{The k-means clustering}

\label{sec:kmeans}

The k-means clustering is a simplification of the Gaussian mixture model
\cite{Press:1058313}. Tracks do not get assigned to clusters in a
probabilistic way but they can be attached to one and only one of them. In the
expectation step the tracks are assigned to the cluster $k$ which has the
closest mean $\mu_k$. In the minimization step the means $\mu_k$ are
re-estimated using the averaged $z$-coordinate of tracks assigned to cluster
$k$. The process is stopped if the expectation step does not change the
assignment of any track. The k-means clustering is simpler than the Gaussian
mixture model, it is fast and converges rapidly, with somewhat reduced
performance. For performance plots of this method see Sec.~\ref{sec:result}
(labeled with fPNN+kMeans).

\subsection{Estimating the number of primary vertices}

Since the number of primary vertices $K$ is not known in advance, it has to be
determined from data. Starting with $K=1$, the value of $\chi^2(K)$ as function
of $K$ can be examined. If adding a new vertex decreases $\chi^2$ by some
substantial amount, the addition can be regarded successful, otherwise the
vertices found in the previous step should be retained.

If there were indeed $K$ real vertices in the bunch crossing, the expected
value of $\widehat{\chi^2}(K)$ assuming distinct vertices, that is,
non-overlapping clusters of tracks, is
\begin{equation}
 \widehat{\chi^2}(K) \approx \sum_n \frac{(z_n - \widehat{z}_k)^2}{\sigma_n^2}
  - 2 \sum_k n_k \log \frac{n_k}{N}
 \label{eq:expected}
\end{equation}

\noindent where $n_k$ are the number of tracks associated to the vertices as
result of the optimization described in Sections~\ref{sec:agglo} and
\ref{sec:class}.  $\widehat{\chi^2}$ has a shifted chi-squared distribution
with $N$ degrees of freedom
\begin{gather*}
 P(\widehat{\chi^2}|K) = f(\widehat{\chi^2} - \lambda; N) \\
\intertext{where the shift is}
 \lambda = 2 \left(N \log N - \sum_k n_k \log n_k\right).
\end{gather*}

The task is to compare the merit values based on individual track probabilities
(Eq.~\eqref{eq:loglikelihood}) and the above calculated expected values
(Eq.~\eqref{eq:expected}). If $\chi^2(K)$ is compatible with
$\widehat{\chi^2}(K)$ with some confidence, $K$ can be regarded as a good
estimate of the number of interaction vertices. A sensible stopping condition
is
\begin{equation*}
 P(\chi^2 - \lambda; N) > 10^{-3}.
\end{equation*}

\subsection{Use of priors}

If the shape of the interaction region or the interaction rate is unknown, we
have no additional information, the priors are constant. Otherwise the
$z$-distribution of vertices $P(\hat{z}_k)$ and the number distribution of
primary vertices $P(K | K \ge 1)$ can be incorporated into the optimization.
Their distributions are well described by a Gaussian and a Poissonian,
respectively:
\begin{gather*}
 P(\hat{z}_k) = \frac{1}{\sigma_{IR} \sqrt{2\pi}}
  \exp\left[-\frac{(\hat{z}_k - z_{IR})^2}{2\sigma_{IR}^2}\right] \\
 P(K | K \ge 1) = \frac{\mu^K e^{-\mu}}{K!} \frac{1}{1 - e^{-\mu}}
\end{gather*}

\noindent
where $z_{IR}$ and $\sigma_{IR}$ are the mean and the standard deviation of the
longitudinal coordinate of the interaction region, $\mu$ is the average number
of interaction vertices per bunch crossing.
The contributions of the priors corresponding to the interaction region ($IR$)
and the number of reconstructible inelastic interactions ($int$) to $\chi^2$ are
\begin{align*}
 \Delta\chi^2_{IR}  &= \sum_k \frac{(z_k-z_{IR})^2}{\sigma_{IR}^2} \\
 \Delta\chi^2_{int} &= 2 \left[ \mu + \log(1 - e^{-\mu})
                                + \log K! - K \log \mu \right].
\end{align*}

\begin{figure}
 
 \begin{center}
  \input{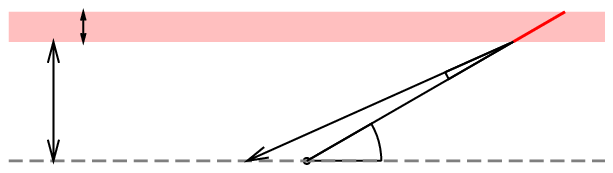}
 \end{center}

 \caption{The effect of multiple scattering. The first silicon layer is shown
by the thick horizontal band (not on scale), the position of the primary vertex
is indicated by $V$. For details see the text in
Sec.~\ref{sec:multipleScattering}.}

 \label{fig:multipleScattering}

\end{figure}

\section{Simulation}

\label{sec:simulation}

Inelastic proton-proton collisions were simulated using the Pythia
event generator \cite{Sjostrand:2006za} at a center-of-mass energy of 10~TeV
(including single-, double-diffractive and non-diffractive events). In order to
take into account the limited acceptance of central detectors, only those
primary charged particles were used for vertex finding which had $|\eta| < 2.5$
and $p_T >$ 0.1 GeV/$c$. The shape of the interaction region in $z$ was
approximated by a Gaussian with 5 cm standard deviation. The distribution of
pseudo-rapidity, $p_T$ and event-by-event multiplicity are shown in
Fig.~\ref{fig:sim}.

At least two compatible tracks are required to form an interaction vertex
candidate ($n_k \ge 2$) which eliminates most of the background coming from low
$p_T$ spiralling particles looping in the strong magnetic field, decay products
of weakly-decaying long lived resonances ($\PKzS$, $\PgL$ and $\PagL$), $\Pgg$
conversions, and particles from secondary interactions.

\subsection{Resolution of z position}

The resolution of $z$ is governed by the local position resolution of
the closest situated (pixel) silicon detectors and the effect of multiple Coulomb scattering.
\begin{gather*}
 \sigma_z^2 = \sigma_{pos}^2 + \sigma_{ms}^2.
\end{gather*}

\label{sec:multipleScattering}

\noindent
In this study $\sigma_{pos} =$ 50 $\mu$m is assumed. The expected standard
deviation of $z$ due to multiple scattering can be written in the form
\begin{multline*}
 \sigma_{ms} = \frac{r}{\sin^2\theta} \theta_0 =
                     \frac{r}{\sin^2\theta}
                \frac{13.6~\mathrm{MeV}}{\beta p c}
                 \sqrt{\frac{x}{X_0 \sin\theta}} \approx \\ \approx
 \frac{100~\mathrm{\mu m}}{p_T c} \cosh^{3/2}\eta
\end{multline*}

\noindent where $\theta$ is the polar angle, $\theta_0$ is the standard
deviation of the multiple scattering angle, $r$ is  the radial distance of the
first pixel layer, $x/X_0$ is the thickness of the layer in radiation length
units.  For detailed derivation see Fig.~\ref{fig:multipleScattering}. Here
the values $r =$ 4 cm, $x/X_0 =$ 3\% were used.

\section{Results}

\label{sec:result}

The performance of several discussed vertex finding methods was compared using
$10^5$ inelastic events. Only those simulated and reconstructed vertices were
taken into account which had at least two tracks.

\begin{figure}

 \begin{center}
  \input{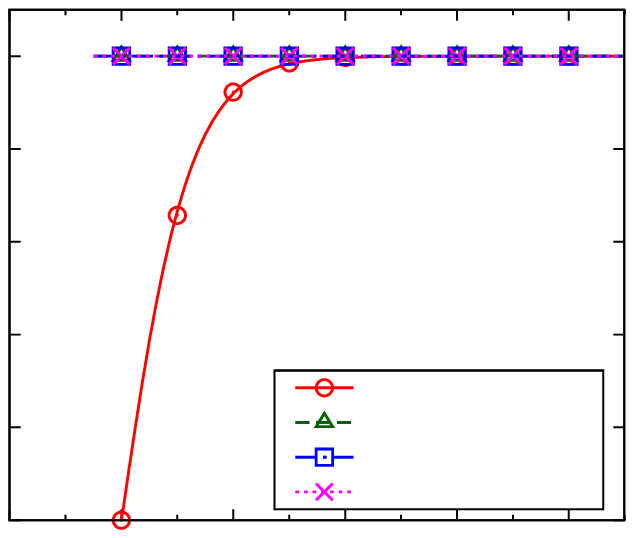}
 \end{center}

 \caption{Efficiency of vertex finding for single events, as a function of
track multiplicity, for several discussed vertex finding methods. The line is
a binomial fit to the values corresponding to the standard algorithm.}

 \label{fig:efficiency}

\end{figure}

The efficiency of vertex finding for single events, as a function of track
multiplicity, for several discussed vertex finding methods is shown in
Fig.~\ref{fig:efficiency}. In case of the standard algorithm the values were
fitted with a binomial distribution, giving a true probability $p_{standard} =
0.87$.

\begin{figure*}
 
 \begin{center}
  \input{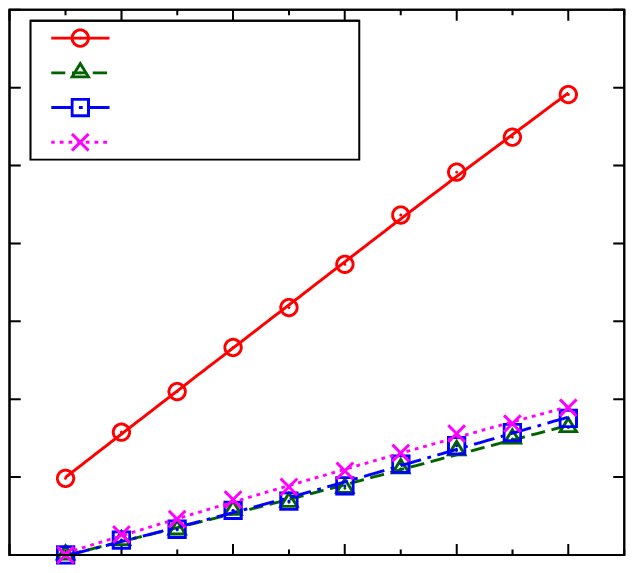}
  \input{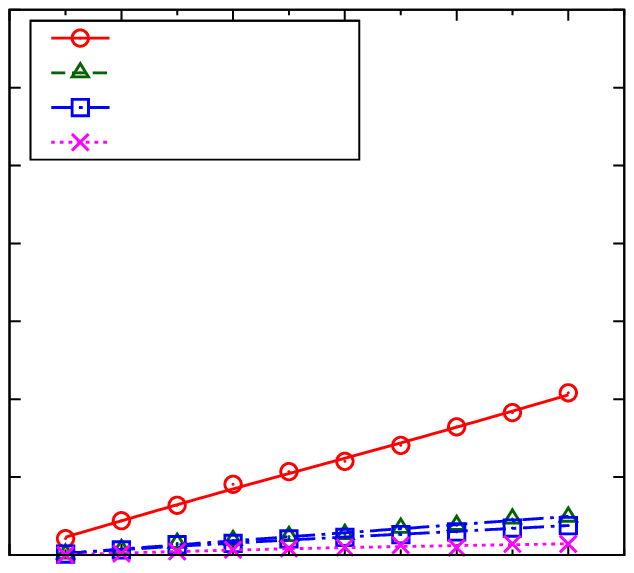}
 \end{center}
 
 \vspace{-0.4in}
 
 \begin{center}
  \input{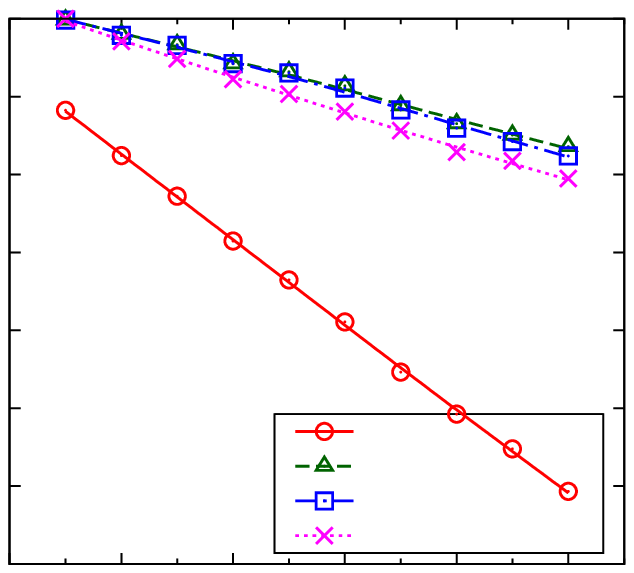}
  \input{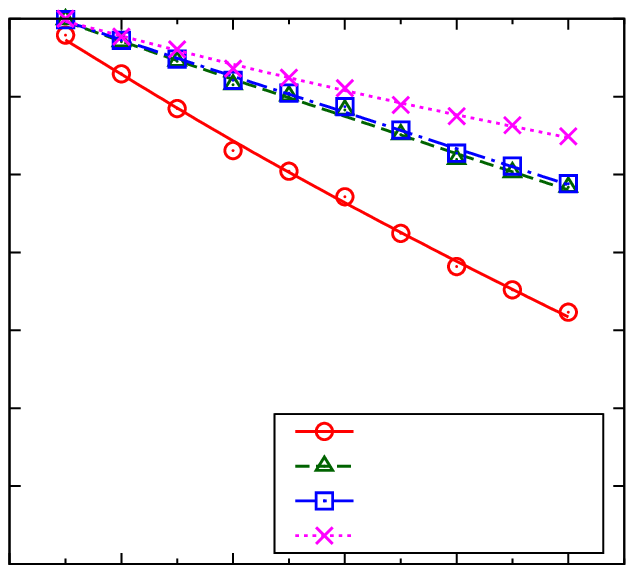}
 \end{center}
 
 \vspace{-0.4in}
 
 \begin{center}
  \input{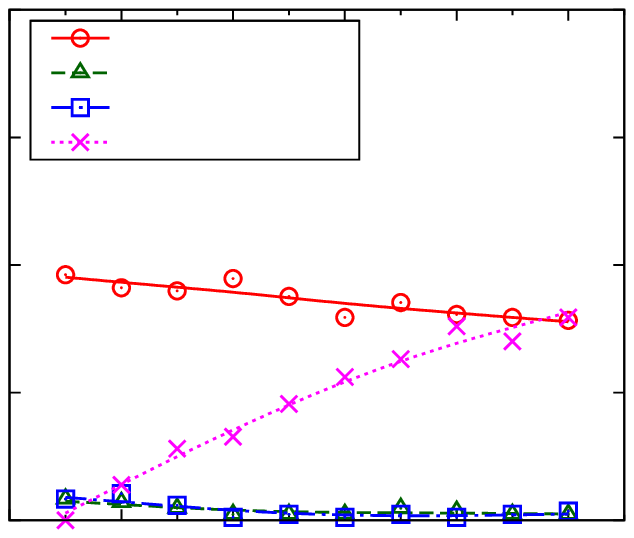}
  \input{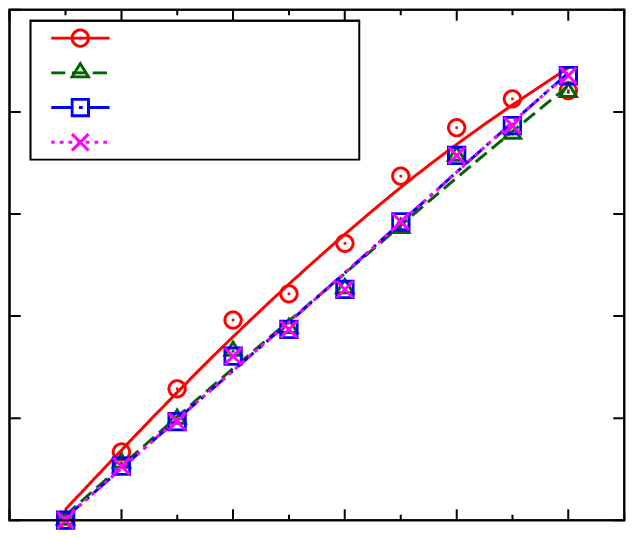}
 \end{center}

 \caption{Comparison of performance for several discussed vertex finding
methods. The left column deals with simulated vertices, by showing those which
have zero (lost vertex), one (singly reconstructed) and more than one (split
vertex) associated reconstructed vertices. The right column takes into account
reconstructed vertices, by looking at those which have zero (fake vertex), one
(correctly found) or more than one (merged vertex) associated simulated
vertices. The lines are drawn to guide the eye.}

 \label{fig:performance}

\end{figure*}

\begin{figure*}

 \begin{center}
  \input{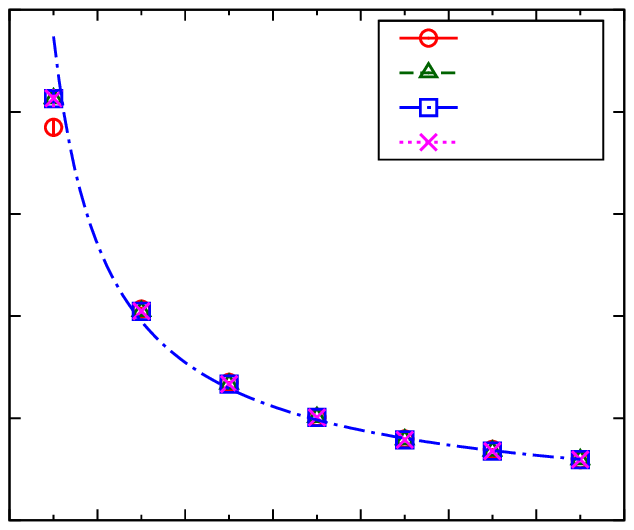}
  \input{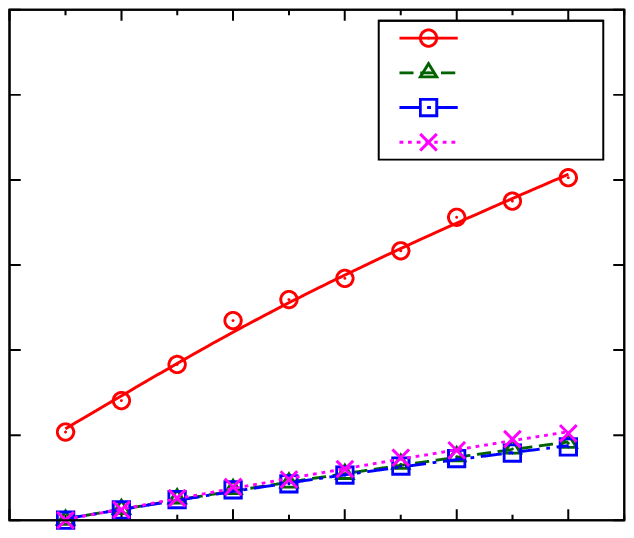}
 \end{center}

 \caption{Left: Resolution of the longitudinal coordinate of the reconstructed
primary vertex as a function of track multiplicity in case of single events,
for several discussed vertex finding methods. The line is a power-law fit to
the values corresponding to the Gaussian mixture algorithm, its exponent
$\alpha_{GaussM}$ is indicated in the figure. Right: Average fraction of lost
vertex tracks as a function of vertex multiplicity, for several discussed
vertex finding methods. The lines are drawn to guide the eye.}

 \label{fig:resolutionLost}

\end{figure*}

The result as a
function of vertex multiplicity in a bunch crossing is shown in
Fig.~\ref{fig:performance}.
The left column deals with simulated vertices by showing the fraction lost,
singly reconstructed and split ones, from top to bottom.
The fraction of lost vertices increases approximately linearly with vertex
multiplicity $K_{sim}$, the improved methods have about 3 times fewer lost
vertices. The rate of split vertices is flat. While it is at 1\% for the
standard method, it drops to 0.1\% for the proposed ones.
The right column takes into account reconstructed vertices by looking at the
fraction of fake, correctly found and merged ones, from top to bottom.  The
fraction of fake vertices is again increases approximately linearly with
reconstructed vertex multiplicity $K_{rec}$, the improved methods have about 5
times fewer fakes. The rate of merged vertices has a linear behavior which is
similar for both standard and improved methods. It can be understood since with
increasing multiplicity the vertices get closer and it is more and more
difficult to separate them.
It it clear that the improved methods have better
performance in all examined variables. The fPNN
search, the k-means clustering and the Gaussian mixture model all provide very
similar values.

The resolution of the longitudinal coordinate of the reconstructed primary
vertex as a function of track multiplicity is shown in
Fig.~\ref{fig:resolutionLost}-left. It is practically independent of the
vertexing method used and scales as $N^{-0.81}$. (The small difference for
$N$=5 is due to the high efficiency of the proposed methods at very low track
multiplicity.) A measure of efficiency, the average fraction of lost vertex
tracks, as a function of vertex multiplicity is plotted in
Fig.~\ref{fig:resolutionLost}-right. It shows that even in case of a single
vertex, the standard method loses about 10\% of the tracks, while the improved
ones keep all of them. The loss increases approximately linearly with $K_{sim}$
multiplicity, the improved methods have about 4 times fewer lost vertex tracks.

\begin{figure*}

 \begin{center}
  \input{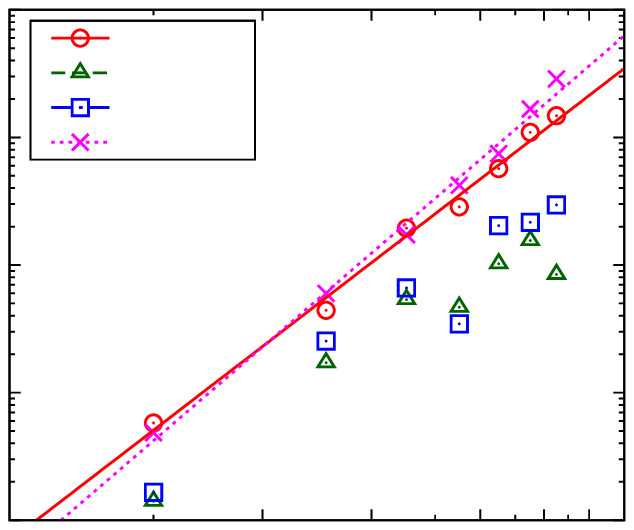}
  \input{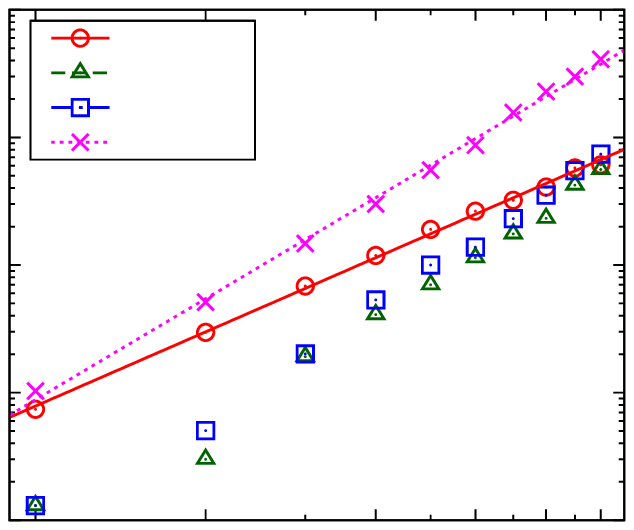}
 \end{center}

 \caption{Processing time per bunch crossing for several discussed methods and
processes on a 1.6~GHz CPU. Left: single events, as a function of track
multiplicity. Right: multiple events, as a function of vertex multiplicity.
The exponents of the power-law fits for the standard ($\alpha_{standard}$) and
fPNN methods ($\alpha_{fPNN}$) are indicated in the figure.}

 \label{fig:timing}

\end{figure*}

\subsection{Timing}

For online applications and fast event reconstruction it is important to
control the timing of vertex finding. Processing times per bunch crossing for
several discussed methods and processes are shown in Fig.~\ref{fig:timing}.
Measured values on a 1.6~GHz CPU are given for the standard method, the fPNN
clustering, as well as for the k-means and Gaussian
mixture models. In case of the improved sequence, both for single events and
pile-up, the largest contribution to timing comes from the clustering phase.
For single events (Fig.~\ref{fig:timing}-left), the required time per
interaction has a power-law dependence on  the track multiplicity, the
exponents being similar for the standard ($\alpha_{standard} =$ 2.2) and
the improved method ($\alpha_{fPNN} =$ 2.4). The processing times are
essentially the same.
In case of multiple events per bunch crossing (Fig.~\ref{fig:timing}-right),
pile-up, the power-law scaling is steeper for the improved method
($\alpha_{fPNN} =$ 2.6) with respect to the standard one
($\alpha_{standard} =$ 1.9). This amounts to a 5 times slower processing
in case of 10 simulated vertices, where the total time per bunch crossing is
4~ms, but that is still acceptable.

\subsection{Sensitivity}

 The stability of the results was also tested, since the performance of the
proposed method can be sensitive to

\begin{itemize}

 \item the amount of background tracks compatible with the beam-line (see
Sec.~\ref{sec:simulation}). In the test their number was set to 2\% of
the primary multiplicity.

 \item systematic shift in the estimated standard deviation of the $z$ value of
tracks. In the test $\sigma_z$ was uniformly increased and decreased by 10\%.

 \item random shifts in the estimated standard deviation of the $z$ value of
tracks. In the test $\sigma_z$ was track-by-track varied according to a
normal distribution with 10\% standard deviation.

\end{itemize}

\begin{figure*}

 \begin{center}
  \input{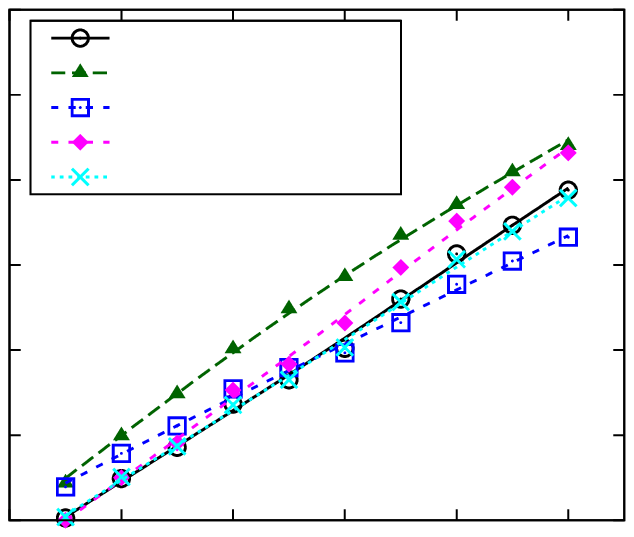}
  \input{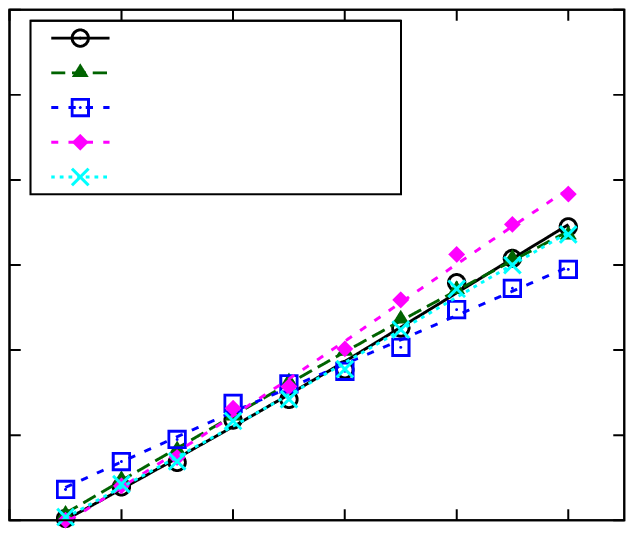}
 \end{center}

 \caption{Sensitivity of the Gaussian mixture method to background tracks.
Comparison of the baseline, underestimated or overestimated $\sigma_z$, and
uncertain $\sigma_z$ scenarios as a function of vertex multiplicity, based on
the merit value $X^2$, for two $n_{min}$ settings (2 or 3).}

 \label{fig:sensitivity}

\end{figure*}

The comparisons based on the merit value $X^2$ are shown in
Fig.~\ref{fig:sensitivity}-left. In case of the addition of background tracks
the slightly worse performance originates from the increase of fake vertex rate.
The underestimation of $\sigma_z$ leads to increased fake and split vertices
for lower vertex multiplicity. The overestimation of $\sigma_z$ gives higher
rate of lost and merged vertices for higher vertex multiplicity. The effects
can be compensated by increasing the parameter $n_{min}$ to 3
(Fig.~\ref{fig:sensitivity}-right). Random shifts in $\sigma_z$ have
practically no effect on the performance.

\section{Conclusions}

It was shown that finding interaction vertices for collider detectors can be
improved by using advanced clustering and classification methods, such as
agglomerative clustering,
followed
by Gaussian mixture model or k-means clustering. The improvement is present
already for single events but it is most pronounced for pile-up. The better
performance for minimum bias proton-proton collisions means less lost vertices
(one third), very few split, and less fake vertices (one fifth). The number of
lost vertex tracks is decreased as well (one fourth). The scaling of the timing
and the sensitivity of the proposed method were studied.

\section*{Acknowledgements}

The author wishes to thank to Wolfgang Adam, Kriszti\'an Krajcz\'ar for helpful
discussions. This work was supported by the Hungarian Scientific Research Fund
and the National Office for Research and Technology (K 48898, H07-B 74296).

\bibliographystyle{elsarticle-num}
\bibliography{vertexFinder}

\end{document}